\documentclass[10pt,twoside]{hsqcd}
 \usepackage{epsf,amsmath}
  \usepackage[dvips,final]{graphicx}
   \usepackage{amssymb}
    \usepackage{amsmath}
     \usepackage{pifont}
\def\MS{$\overline{\text{MS}\vphantom{^1}}${ }}
\begin{document}
\title{{\hfill RUB-TPII-01/06}\\ [1cm]Fractional Analytic 
        Perturbation Theory in QCD and Exclusive Reactions\thanks{Invited
        plenary talk presented by the first author at \textit{Hadron
        Structure and QCD: from Low to High Energies}, St. Petersburg,
        Russia, 20-24 September 2005.}}
\author{N.~G.~Stefanis$^{1,a}$, A.~P.~Bakulev$^{2,b}$, 
        A.~I.~Karanikas$^{3,c}$, 
        S.~V.~Mikhailov$^{2,d}$,   \\
$^1$Institut f\"ur Theoretische Physik II,
    Ruhr-Universit\"at Bochum, \\
    D-44780 Bochum, Germany    \\
$^a$Email: stefanis@tp2.ruhr-uni-bochum.de\\
$^2$Bogoliubov Laboratory of Theoretical Physics, JINR,
    141980 Dubna, Russia       \\
$^b$Email: bakulev@thsun1.jinr.ru \\
$^d$Email: mikhs@thsun1.jinr.ru \\
$^3$University of Athens, Department of Physics, Nuclear and Particle\\ 
    Physics Section, Panepistimiopolis, GR-15771 Athens, Greece\\
$^c$Email: akaran@phys.uoa.gr}
\maketitle

\begin{abstract}
\noindent Different ``analytization'' procedures for the factorized 
pion form factor are discussed in comparison with the standard QCD 
perturbation theory at NLO.
It is argued that demanding the analyticity of the exclusive 
amplitude as a \emph{whole}, entails insensitivity of the results on 
all scheme and scale-setting parameters, including the factorization 
scale. 
This enables us to develop an approach of optimized perturbation 
theory within the \MS scheme and to generalize the Analytic 
Perturbation Theory to non-integer (fractional) powers of the strong 
running coupling in the complex $Q^2$ plane.
\end{abstract}



\markboth{\large \sl N.~G.~Stefanis et al.  \hspace*{2cm} HSQCD 2005}
{\large \sl \hspace*{1cm} Fractional Analytic Perturbation Theory 
                          and Exclusive Reactions}

\section{Introduction}
\label{sec:intro}
Quantum field theories are plagued with infinities.
While those infinities related to the ultraviolet (UV) properties of
the theory can be explained away by means of renormalization,
singularities in the infrared (IR) are more subtle to handle.
QCD is a renormalizable theory and possesses asymptotic freedom.
However, the strong running coupling develops 
at $Q^2=\Lambda_{\rm QCD}^2$ an artificial singularity, 
termed (in one-loop) the Landau pole, that prevents the application 
of perturbative QCD in the low-momentum spacelike region.
Moreover, hadronic quantities calculated at the partonic level are
expressed in terms of a power-series expansion in the running
coupling and are strongly affected by this IR divergence---in
particular in that momentum region accessible to experiment.

Truncating this expansion, the result depends on the particular
choice of the renormalization scheme and scale, though, on account of
the renormalizability of QCD, all-order expressions in different 
schemes would be the same.
Truncated series are numerically \emph{not} equal and hence one has
to design a scheme and specify a renormalization scale, which minimize
the contribution of the discarded terms.
In addition, employing a convolution approach to isolate the
short-distance part of the process in question, causes beyond leading
order (LO) of perturbative QCD (pQCD) a dependence of the result on the 
factorization scale.
Much has been written about these problems, but until recently no
satisfactory answers were provided.

\section{Infrared-finite QCD coupling and ``analytization'' approaches}
\label{sec:FAPT}

The situation improved dramatically during the last few years with the
development of Analytic Perturbation Theory (APT) by Shirkov, 
Solovtsov, Milton, and Solovtsova (SSMS) 
\cite{SS97,MS97}.
In this scheme the running coupling and its powers are 
replaced by singularity-free expressions in the spacelike 
regime using renormalization-group invariance and causality.
The conversion to analytic images of the QCD coupling,
$a=b_{0}\alpha_{s}/4\pi$,  at the $l$-loop order,
${\cal A}_{n}^{(l)}(L)=[a_{(l)}^{m}(Q^2)]_{\rm an}$,  
is based on the dispersion relation
\begin{equation}
  \left[f(Q^2)\right]_{\rm an}
   = \int_0^{\infty}\!
      \frac{\rho_f(\sigma)}
         {\sigma+Q^2-i\epsilon}\,
       d\sigma
\label{eq:disp-rel}
\end{equation}
with the spectral density 
$\rho_{f}(\sigma)={\bf Im}\,\big[f(-\sigma)\big]/\pi$.
The coupling (and its powers) can be analytically continued to the 
timelike ($s$-channel) region:
\begin{eqnarray}
  \{ a^n(Q^2)\} \; \to \!\!\! &\left\{
       \begin{array}{rl}
       & \!\!\!\!\!\!\!\!\{{\cal A}_{n}(L)\}_{n\in\mathbb{N}} 
       \quad\;\,
  L=\ln Q^2/\Lambda^2 \nonumber \quad (-q^2=Q^2), 
  \displaystyle ~~{\cal A}_{1}(L)=\frac{1}{L} - \frac{1}{(e^{L}-1)}
   \nonumber \\ 
       & \!\!\!\!\!\!\!\!\{{\mathfrak A}_{n}(L_{s})\}_{n\in\mathbb{N}} 
       \quad 
  L=\ln s/\Lambda^2 \quad\;\;\; (q^2=s)\,  
       \end{array} \, .
 \right. 
 \label{eq:logs}
 \end{eqnarray}
But pQCD higher-order calculations (or evolution factors) entail 
expressions like 
$
 \left[a(L)\right]^\nu,
$
$
 \left[a(L)\right]^\nu\ln^{m}\!\left[a(L)\right],
$
$ 
a^\nu L^m, e^{-a(L)f(x)}
$
that are not covered by the SSMS ``analytization'' scheme.
Such terms contribute to the spectral density
and their analytic images are inevitably required. 
It was shown \cite{BMS05} that, using the Laplace transformation in 
conjunction with dispersion relations, closed-form expressions for 
the analytic images of the running coupling powers, $a_{s}^{\nu}$, 
for any \emph{fractional} (real) power $\nu$ can be derived.
In the spacelike region, these images can be expressed in terms 
of the reduced transcendental Lerch function $F(z,\nu)$ \cite{BE53} 
(compare with ${\cal A}_{1}(L)$ in Eq.\ (\ref{eq:logs})):
\begin{equation}
  {\cal A}_{\nu}(L) 
      = \frac{1}{L^\nu} 
      - \frac{F(e^{-L},1-\nu)}{\Gamma(\nu)}\, ,
\label{eq:lerch}
\end{equation}
where the first term corresponds to pQCD and the second one 
is entailed by the pole remover.
This function is an entire function in the index $\nu$ and has the 
properties 
${\cal A}_{0}(L)=1$, 
${\cal A}_{-m}(L)=L^{m}$ for $m\in\mathbb{N}$, and 
${\cal A}_{m}(\pm\infty)=0$ for $m\geq 2$, $m\in\mathbb{N}$, while 
for $|L|<2\pi$, it reads 
$
 {\cal A }_{\nu}(L) 
= 
 -\left[1/\Gamma(\nu)\right]\sum_{r=0}^{\infty}\zeta(1-\nu-r)
                            \left[(-L)^{r}/r!\right]. 
$ 
In the timelike region, these images are completely determined 
by elementary functions \cite{BMS05}:
\begin{equation}
  {\mathfrak A}_{\nu}(L) 
    = \frac{\sin\left[(\nu -1)\arccos\left(L/\sqrt{\pi^2+L^2}\right)
                \right]}
      {\pi(\nu -1) \left(\pi^2+L^2\right)^{(\nu-1)/2}} \, . 
\label{eq:gothic}
\end{equation}
We also defined the index derivative, needed to describe terms 
that contain products of coupling powers and logarithms (see Table 1).  
All this implements the Karanikas-Stefanis (KS) analyticity requirement 
\cite{KS01} imposed on hadronic quantities in QCD at the amplitude 
level and generalizes APT to Fractional Analytic Perturbation Theory 
(FAPT) \cite{BMS05}.
 \begin{table}[h]
 \caption{Comparison of pQCD and FAPT}
  \begin{tabular}{|c|c|c|c|}\hline
  ~~~~~~Theory~~~~~~ 
           & ~~~~~~~~PT~~~~~~~~ 
                     & ~~~~~APT~~~~~  
                               & ~~~~~FAPT~~~~~         \\ 
  \hline\hline
  Space  & $\big\{a^\nu\big\}_{\nu\in\mathbb{R}}\vphantom{^{\big|}_{\big|}}$
         & $\big\{{\cal A}_m\big\}_{m\in\mathbb{N}}$
         & $\big\{{\cal A}_\nu\big\}_{\nu\in\mathbb{R}}$
                                                        \\ 
  Series expansion
         & $\sum\limits_{m}f_m\,a^m(L)\vphantom{^{\big|}_{\big|}}$
         & $\sum\limits_{m}f_m\,{\cal A}_m(L)$
         & $\sum\limits_{m}f_m\,{\cal A}_m(L)$
                                                        \\ 
  Inverse powers
         & $\left[a(L)\right]^{-m}\vphantom{^{\big|}_{\big|}}$
         & ~{---}~
         & ${\cal A}_{-m}(L)=L^m$
                                                        \\ 
  Multiplication
         & $a^{\mu} a^{\nu}= a^{\mu+\nu}\vphantom{^{\big|}_{\big|}}$
         & ~{---}~
         & ~{---}~                                      \\ 
  Index derivative
         & $a^{\nu} \ln^{k}a\vphantom{^{\big|}_{\big|}}$
         & ~{---}~
         & ${\cal D}^{k}{\cal A}_\nu\equiv \frac{d^{k}}{d\nu^{k}}{\cal A}_\nu
          =\left[a^{\nu}\ln^{k}(a)\right]_{\rm an}$
                                                        \\ 
  \hline
\end{tabular}
\end{table}

\section{Electromagnetic pion form factor in FAPT}
\label{sec:pionFF}
  
The ``analytization'' concept has been applied to the factorized part 
of the pion form factor in next-to-leading order (NLO) pQCD 
\cite{MNP99}, including also Sudakov effects due to soft-gluon 
emission, in \cite{SSK99}.
Subsequently, the power-series expansion of $F_{\pi}^{\rm Fact}(Q^2)$ 
in terms of the analytic coupling (``naive analytization'') at NLO was 
traded in \cite{BPSS04} in favor of a non-power-series (functional) 
expansion in terms of analytic images of the coupling and its powers, 
with the coefficients $d_m$ being numbers in the \MS scheme,
$
 \sum\limits_{m}d_m a_s^m(Q^2)
\Rightarrow
 \sum\limits_{m}d_m {\cal A}_{m}(Q^2)
$ 
(``maximal analytization''), taking into account NLO ERBL\ evolution 
and accounting for heavy-quark threshold effects.
This NLO pQCD treatment has provided predictions for 
\begin{equation}
  F_{\pi}^{\rm Fact}(Q^2)
=
  \varphi_\pi(x,\mu_F^2)\otimes
  T^{\rm NLO}_{\rm H}\left(x,y,Q^2;\mu_{F}^2,\mu_{R}^2\right)
                        \otimes
  \varphi_\pi(y,\mu_F^2)
\label{eq:F-fact}
\end{equation}
that are stable against changes of the renormalization scheme and
associated scale settings.
Here all nonperturbative information is encapsulated in the leading 
twist-2 pion distribution amplitude (DA) :
\begin{eqnarray}
 \varphi_\pi(x,\mu^2)
  = 6 x (1-x)
     \left[ 1
          + a_2(\mu^2) \, C_2^{3/2}(2 x -1)
          + a_4(\mu^2) \, C_4^{3/2}(2 x -1)
          + \ldots
     \right]
\label{eq:phi024mu0}
\end{eqnarray}
in terms of the Gegenbauer coefficients $a_n$.
Below, predictions are shown for a model DA, derived in \cite{BMS01}
by means of nonlocal QCD sum rules ($\mu^2 \approx 1$~GeV${}^{2}$).  

More recently \cite{BKS05}, the ``analytization'' of the 
electromagnetic pion form factor was performed within FAPT at 
two-loop order including into the ``analytization'' process also 
logarithms of the factorization scale that have non-zero spectral 
density.
Recall (cf. Eq.\ (\ref{eq:lerch})) that the pole remover does not 
contribute to the spectral density; the discontinuity is determined 
solely by the term $1/L$.
This leads to the following expression for the hard-scattering 
amplitude---with the renormalization scale set equal to 
$\mu_{\rm R}^{2}=\lambda_{\rm R}Q^2$:
\begin{eqnarray}
 && \left[Q^2 T_{\rm H}\left(x,y,Q^2;\mu^{2}_{\rm F},\lambda_{\rm R} Q^2
                       \right)
   \right]_{\rm KS}^{\rm an}
  = {\cal A}_{1}^{(2)}\left(\lambda_{\rm R} Q^2
                      \right)\, t_{\rm H}^{(0)}(x,y)
 ~~~~~~~~~~~~~~~~~~~~~~~~~~~~~~~~~~~~
 \nonumber\\ 
&& ~~~~~~~~~~ +\ \frac{{\cal A}_{2}^{(2)}\left(\lambda_{\rm R} Q^2
                                         \right)}{4\pi}\,
       \left[b_0\,t_{\rm H}^{(1,\beta)}(x,y;\lambda_{\rm R})
           + t_{\rm H}^{({\rm FG})}(x,y)
           + C_{\rm F}\,
             t_{{\rm H},2}^{(1,{\rm F})}
             \left(x,y;\frac{\mu^{2}_{\rm F}}{Q^2}\right)
       \right]
       \nonumber\\
&& ~~~~~~~~~~ +\ \frac{\Delta_{2}^{(2)}
  \left(\lambda_{\rm R} Q^2\right)}{4\pi}\,
      \left[C_{\rm F}\, t_{\rm H}^{(0)}(x,y)  \,
             \left(6 + 2 \ln(\bar{x}\bar{y})\right)
      \right]\, ,\!\!\!\!\!\!\!\!\!\!\!\!\!\!\!
\label{eq:TH-KS-6}
\end{eqnarray}
where $\bar{x}\equiv 1-x$. 
The deviation from its counterpart within the \emph{maximal} 
``analytization'' procedure of \cite{BPSS04} is encoded in the term 
\cite{BKS05}
\begin{eqnarray}\label{eq:delta2-2}
 \Delta_{2}^{(2)}\left(Q^2\right)
  &\equiv&
   {\cal L}_{2}^{(2)}\left(Q^2\right)
    - {\cal A}_{2}^{(2)}\left(Q^2\right)\,\ln\left[Q^2/\Lambda^2\right]
\end{eqnarray}
with
\begin{eqnarray}\label{eq:Log_Alpha_2_KS}
 {\cal L}_{2}^{(2)}\left(Q^2\right)
  &\equiv&
   \left[\left(\alpha_{s}^{(2)}\left(Q^2\right)\right)^2
         \ln\left(\frac{Q^2}{\Lambda^2}\right)
   \right]_{\rm KS}^{\rm an}
  =  \frac{4\pi}{b_0}\,
     \left[\frac{\left(\alpha_{s}^{(2)}\left(Q^2\right)\right)^2}
                 {\alpha_{s}^{(1)}\left(Q^2\right)}
      \right]_{\rm KS}^{\rm an}\, .
\end{eqnarray}
Here $\ln(Q^2/\mu^{2}_{\rm F})
=
 \ln (\lambda_{\rm R} Q^2/\Lambda^2) -
 \ln (\lambda_{\rm R}\mu^{2}_{\rm F}/\Lambda^2)$.
Performing the ``analytization'' one finds
\begin{eqnarray}\label{eq:Log_Alpha_2_BMKS}
  {\cal L}_{2}^{(2)}\left(Q^2\right)
   = \frac{4\pi}{b_0}\,
      \left[{\cal A}_{1}^{(2)}\left(Q^2\right)
      + c_1\,\frac{4\pi}{b_0}\,f_{\cal L}\left(Q^2\right)
      \right]\, ,
\end{eqnarray}
where
\begin{eqnarray}\label{eq:f_MS}
  f_{\cal L}\left(Q^2\right)
   = \sum_{n\geq0}
      \left[\psi(2)\zeta(-n-1)-\frac{d\zeta(-n-1)}{dn}\right]\,
       \frac{\left[-\ln\left(Q^2/\Lambda^2\right)
             \right]^n}{\Gamma(n+1)}
\end{eqnarray}
and $\zeta(z)$ is the Riemann zeta-function.

\begin{figure}[th]
 \centerline{\includegraphics[width=0.44\textwidth]{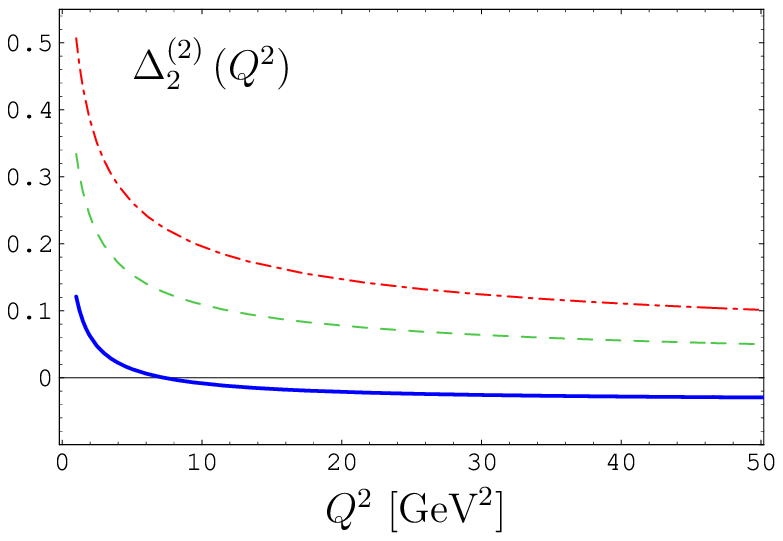}~~~%
  \includegraphics[width=0.44\textwidth]{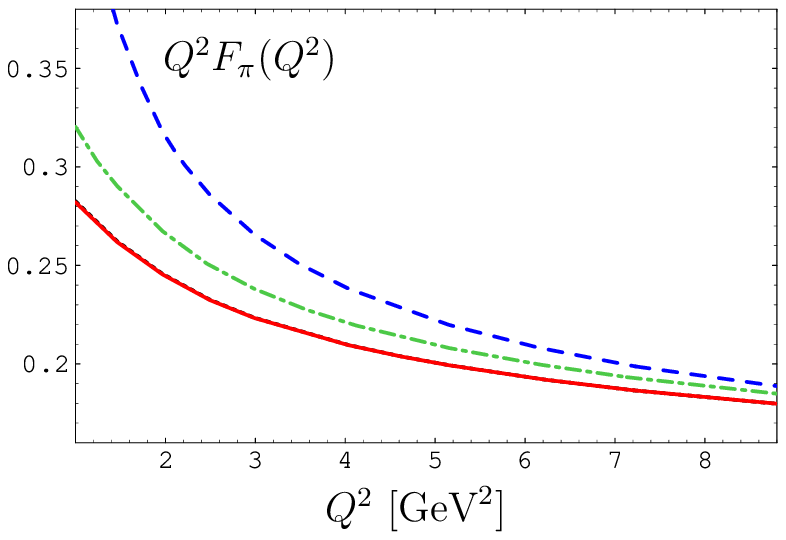}}
   \caption[*]{Left: Difference corresponding to Eqs.\ 
    (\protect\ref{eq:Log_Alpha_2_BMKS}) and (\protect\ref{eq:f_MS}) 
    between ``maximal'' and KS 
    ``analytization'' in $Q^2T_{\rm H}^{\rm NLO}$ (solid line).
    The other curves are approximations explained in 
    \protect\cite{BKS05}.
    Right: Results for the factorized pion form factor, scaled with 
    $Q^2$, and setting 
    $\mu_{\rm R}^2=Q^2$, $\mu_{\rm F}^2=5.76$~GeV$^2$ in pQCD (dashed 
    line), dash-dotted line---\emph{naive} APT; 
    solid line---\emph{maximal} APT.
    $[Q^2F_{\pi}^{\rm Fact}(Q^2)]_{\rm KS}$ almost
    coincides with $[Q^2F_{\pi}^{\rm Fact}(Q^2)]_{\rm max}$ (not 
    shown). 
\label{fig:softFF}}
\end{figure}

The main upshot of the FAPT analysis is that the dependence of the 
prediction for $F_{\pi}^{\rm Fact}(Q^2)$ on all perturbative scheme 
and scale settings is diminished already at NLO.
In addition to the renormalization-scale stability already achieved 
with the ``maximal analytization'' in \cite{BPSS04}, the prediction
now becomes insensitive also to the variation of the factorization 
scale. 
This offers the possibility to calculate perturbatively hadronic
processes in QCD with high theoretical accuracy in a wide range of
momenta from sub-asymptotic values down to a few hundred MeV.
In Fig.\ \ref{fig:softFF}, we show the contribution to 
$Q^2T_{\rm H}^{\rm NLO}$ induced by the KS ``analytization'' (left
panel), whereas the right panel shows the predictions for 
$Q^2F_{\pi}^{\rm Fact}(Q^2)$, using pQCD, \emph{naive} APT,
and \emph{maximal} APT.

\section{Conclusions}
\label{sec:concl}

The ``analytization'' scheme at the amplitude level---technically
realized by means of FAPT---has so far only been used in fully
worked out detail in the calculation of the factorized pion's 
electromagnetic form factor at NLO \cite{BKS05}.
But the concept \cite{KS01} and the developed mathematical apparatus 
\cite{BMS05}, underlying this specific calculation, is not limited to 
that case.
Moreover, the fact that the predictions derived from it show minimal
sensitivity to both the factorization and the renormalization scale, 
and also to the associated scheme-setting procedure---be it within the 
\MS or the $\alpha_V$ scheme, attaches to it fundamental importance.

One of us (NGS) thanks the Deutsche Forschungsgemeinschaft for a 
travel grant. 
This work was supported in part by the Heisenberg--Landau Programme, 
grant 2005 and the Russian Foundation for Fundamental Research, grant 
No.\ 05-01-00992.

\end{document}